\documentclass[nofootinbib]{revtex4}
\usepackage{graphicx}
\usepackage{latexsym}
\usepackage{amsthm,amsmath,amssymb}
\usepackage{mathrsfs}
\usepackage{color}
\def\be{\begin{equation}}
\def\ee{\end{equation}}
\def\bea{\begin{eqnarray}}
\def\eea{\end{eqnarray}}

\begin{document}

\hfill  USTC-ICTS/PCFT-21-38

\title{Gravitational leptogenesis in teleparallel and symmetric teleparallel gravities}

\author{Mingzhe Li}
\author{Yicen Mou}
\author{Haomin Rao}
\author{Dehao Zhao}
\affiliation{Interdisciplinary Center for Theoretical Study, University of Science and Technology of China, Hefei, Anhui 230026, China}
\affiliation{Peng Huanwu Center for Fundamental Theory, Hefei, Anhui 230026, China}

\begin{abstract}
In this paper, we consider the possibilities of generating baryon number asymmetry in thermal equilibrium within the frameworks of teleparallel and symmetric teleparallel gravities. Through the derivative couplings of the torsion scalar or the non-metricity scalar to baryons, the baryon number asymmetry is indeed produced in the radiation dominated epoch. For gravitational baryogenesis mechanisms in these two frameworks, the produced baryon-to-entropy ratio is too small to be consistent with observations. But the gravitational leptogenesis models within both frameworks have the possibilities to interpret the observed baryon-antibaryon asymmetry. 
\end{abstract}

\maketitle

\section{introduction}

All the observations showed that our universe contains an excess of matter over antimatter. The Planck result \cite{Planck:2018vyg} showed that the left baryon-to-photon ratio is $n_B/n_{\gamma}=(6.12\pm 0.04)\times 10^{-10}$  \cite{ParticleDataGroup:2020ssz}. For theoretical discussions, it is more convenient to use the baryon-to-entropy ratio $n_B/s$ to quantify this asymmetry, which is about $n_B/s\simeq 8.7\times 10^{-11}$ from the observational result because $s\simeq 7.04 n_{\gamma}$ at present time. The origin of the baryon number asymmetry remains a big puzzle in cosmology. Conventionally it was argued that this asymmetry is generated dynamically from an initial baryon symmetric phase as long as the following conditions are satisfied \cite{Sakharov:1967dj}: (1) baryon number non-conserving interactions; (2) C and CP violations; (3) a departure from thermal equilibrium. 

However, if the CPT symmetry is violated, the baryon number asymmetry could be generated in thermal equilibrium \cite{Cohen:1987vi}. For instances, in Refs. \cite{Li:2001st,Li:2002wd,DeFelice:2002ir}, an effective interaction 
\be\label{coupling1}
\mathcal{L}_{in}=\frac{c}{M_{\ast}}\partial_{\mu}\phi J^{\mu}_B~,
\ee
between the dynamical dark energy (quintessence) and baryons were introduced, where a dimensionless coupling constant $c$ and a cut-off mass scale $M_{\ast}$ are considered.  As the universe expands, the background evolution of the scalar field $\phi$ breaks the Lorentz as well as the CPT symmetries spontaneously and provides an effective chemical potential for baryons and an opposite one for antibaryons:
\bea
& &\frac{c}{M_{\ast}}\partial_{\mu}\phi J^{\mu}_B \rightarrow \frac{c}{M_{\ast}}\dot\phi \,n_B=\frac{c}{M_{\ast}}\dot\phi (n_b-n_{\bar{b}})~,\nonumber\\
& &\mu_b=\frac{c}{M_{\ast}}\dot\phi =-\mu_{\bar{b}}~.
\eea
This causes a difference between the distribution functions of baryons and antibaryons when they are in thermal equilibrium, so that produces an excess of baryons over antibaryons \cite{Kolb:1990vq}:
\be\label{chemical}
n_B=\frac{g_b T^3}{6}[\frac{\mu_b}{T}+\mathcal{O}(\frac{\mu_b}{T})^3]\simeq c\frac{\dot\phi T^2}{3M_{\ast}}~,
\ee
where $g_b=2$ counts the internal degrees of freedom of the baryon. In terms of the entropy density $s=(2\pi^2/45)g_{\ast s}T^3$, the baryon-to-entropy ratio is written as
\be\label{nbs1}
\frac{n_B}{s}=\frac{15 c}{2\pi^2}\frac{\dot\phi}{g_{\ast s} M_{\ast} T}~.
\ee
The parameter $g_{\ast s}$ counts the total number of relativistic degrees of freedom in the universe and decreases slightly with the cosmic expanding and cooling \cite{Kolb:1990vq}. For the standard model of particle physics, $g_{\ast s}=106.75$ 
in the early time of radiation domination, i.e., the time after reheating but well before the electroweak phase transition. 
Please note that to generate the baryon number asymmetry (\ref{nbs1}), there should be baryon number violating interactions in thermal equilibrium, otherwise the coupling introduced in Eq. (\ref{coupling1}) does not make sense since $\mathcal{L}_{in}\sim \partial_{\mu}\phi J^{\mu}_B\rightarrow -\phi\nabla_{\mu}J^{\mu}_B=0$ after dropping out a total derivative term. In the early time, with $B$-violating processes in thermal equilibrium,
the baryon-to-entropy ratio (\ref{nbs1}) changes as the universe expands. It is so until the temperature of the universe cools to $T_D$ when the $B$-violating interactions decouple from the thermal bath, after that the baryon-to-entropy ratio remains unchanged. So the relic baryon number asymmetry observed today should be
\be
\left. \frac{n_B}{s}\right |_{T_D}=\frac{15 c}{2\pi^2}\left.\frac{\dot\phi}{g_{\ast s} M_{\ast} T}\right |_{T_D}~.
\ee

In such kind of mechanisms, the scalar field $\phi$ in the coupling (\ref{coupling1}) is not necessarily the dark energy, it can be replaced by other cosmic scalar fields. In Ref. \cite{Davoudiasl:2004gf}, the authors proposed 
the following interaction
\be\label{coupling2}
\mathcal{L}_{in}=\frac{1}{M^2_{\ast}}\partial_{\mu}R J^{\mu}_B~,
\ee
between the curvature scalar and baryons.  As a result, the baryon-to-entropy ratio is 
\be
\frac{n_B}{s}\sim \frac{\dot R}{M_{\ast}^2 T}~.
\ee
Because this baryon number asymmetry is generated by the derivative of the curvature which accounts for the gravitation, this type of baryogenesis models are called ``gravitational baryogenesis".
Unfortunately, within the framework of general relativity (GR), the Einstein equation gives: $R=8\pi G T^{\mu}_{\mu}=8\pi G(1-3w)\rho$, so that $R$ and $\dot R$ vanish in the radiation dominated epoch since
$w=1/3$ and no baryon number asymmetry can be produced in this epoch. In order to circumvent this problem, various possibilities to obtain a non-vanishing $\dot R$ have been considered in Ref. \cite{Davoudiasl:2004gf}: significant trace anomaly effects are included in the radiation dominated time $T^{\mu}_{\mu}\neq 0$;
baryon number asymmetry produced in the reheating period in which the universe has $w=0$, and so on. Later the gravitational baryogenesis has been realized under other modifications, for instances, considering the gravity theory different from GR \cite{Shiromizu:2004cb}, or replacing the coupling $\partial_{\mu}R J^{\mu}_B$ with $\partial_{\mu}f(R) J^{\mu}_B$ \cite{Li:2004hh}, where $f(R)$ is a non-linear function of the curvature scalar. 

In this paper, we are interested in the gravitational baryogenesis within the frameworks of teleparallel gravity (TG) \cite{Tele} and symmetric teleparallel gravity (STG) \cite{Nester:1998mp}. These two frameworks provide gravity models in non-Riemannian ways. Both TG and STG  can be equivalent to GR but are formulated in flat spacetime in which the curvature vanishes. In TG model, gravity is attributed to the spacetime torsion, nevertheless in STG model gravity is identified with the non-metricity. Within the GR equivalent TG model and the GR equivalent STG model, we will first consider the baryogenesis induced by derivative couplings to the baryon current. These couplings are similar to Eq. (\ref{coupling2}) except the curvature scalar being replaced by the torsion scalar and the non-metricity scalar respectively. We will show that in these baryogenesis models the produced baryon-to-entropy ratio are too small to be consistent with the observation result. So we will turn to the gravitational leptogenesis in which the torsion scalar and the non-metricity scalar are coupled derivatively to the current of $B-L$. With appropriate cut-off scales, these gravitational leptogenesis models can generate the required baryon number asymmetry. We would like to point out that the gravitational baryogenesis within the framework of TG has been also studied in Refs. \cite{Oikonomou:2016jjh,Bhattacharjee:2020jfk,Azhar:2020coz} in different ways. The gravitational leptogenesis from gravitational waves in models of inflation was proposed in Ref. \cite{Alexander:2004us}. In the following sections we will expand our studies and demonstrations. In both section II and III we will start with brief introductions of the basics of TG and STG. The readers who are not familiar with these two models may learn more details from some reviews on these subjects, e.g., Refs. \cite{Bahamonde:2021gfp}. 

We will in this paper use the convention of most negative signature for the metric. The spacetime tensor indices are represented in Greek letters $\mu,\nu,\sigma,...=0,1,2,3$ and their spatial components are denoted in Latin letters $i,j,k,...=1,2,3$. The tensor indices in the local Minkowski spacetime are represented in capital Latin letters $A, B, C,...=0,1,2,3$ and the corresponding spatial components are denoted in small Latin letters $a, b, c,...=1,2,3$.

\section{Gravitational leptogenesis within teleparallel gravity}

The TG theory can be considered as a constrained metric-affine theory. It is formulated in a spacetime endowed with a metric $g_{\mu\nu}$ and an affine connection ${\hat{\Gamma}^{\rho}}_{~\mu\nu}$, which is constrained by  the vanishment of the curvature and the metric compatibility, 
\be\label{noncurvature1}
 {\hat{R}^{\rho}}_{~\sigma\mu\nu}\equiv \partial_{\mu}{ \hat\Gamma^{\rho}}_{~\nu\sigma}-\partial_{\nu}{\hat{\Gamma}^{\rho}}_{~\mu\sigma}+{\hat{\Gamma}^{\rho}}_{~\mu\alpha}{\hat{\Gamma}^{\alpha}}_{~\nu\sigma}-{\hat{\Gamma}^{\rho}}_{~\nu\alpha}{\hat{\Gamma}^{\alpha}}_{~\mu\sigma}=0~, ~\hat\nabla_{\rho}g_{\mu\nu}=\partial_{\rho}g_{\mu\nu}-{ \hat\Gamma^{\lambda}}_{~\rho\mu}g_{\lambda\nu}-{ \hat\Gamma^{\lambda}}_{~\rho\nu}g_{\mu\lambda}=0~.
 \ee
Without curvature in this theory, the gravity is identified with the spacetime torsion. The torsion tensor is defined as the antisymmetric part of the affine connection: $\mathcal{T}^{\rho}_{~~\mu\nu}=2\hat{\Gamma}^{\rho}_{[\mu\nu]}$.
In terms of the language of tetrad and spin connection and the general relations: $g_{\mu\nu}=\eta_{AB}e^A_{~\mu}e^B_{~\nu}$ and $\hat{\Gamma}^{\mu}_{~\rho\sigma}=e_A^{~\mu}(\partial_{\rho}e^A_{~\sigma}+\omega^A_{~B\rho}e^B_{~\sigma})$, one can find that 
\be
\mathcal{T}^{\rho}_{~~\mu\nu}=2 e_A^{~\rho}(\partial_{[\mu}e^{A}_{~\nu]}+\omega^{A}_{~B[\mu}e^{B}_{~\nu]})~,
\ee
and the spin connection under the constraints of Eq. (\ref{noncurvature1}) can be expressed as,
\be\label{omega}
\omega_{~B \mu}^{A}=\left(\Lambda^{-1}\right)_{~C}^{A} \partial_{\mu} \Lambda_{~B}^{C}~,
\ee
where $\Lambda^{A}_{~B}$ is an element of an arbitrary Lorentz transformation matrix which is position dependent and satisfies the relation $\eta_{AB}\Lambda^A_{~C}\Lambda^B_{~D}=\eta_{CD}$ at any spacetime point. 

The TG model we are most concerned about is the Teleparallel Equivalent of General Relativity (TEGR), in which the action for gravity is 
\bea\label{TGaction}
S_g=\frac{M_p^2}{2}\int d^4x ~{\|e\|} \mathbb{T}~,
\eea
where $M_p=1/\sqrt{8\pi G}$ is the reduced Planck mass, ${ \|e\|}=\sqrt{-g}$ is the determinant of the tetrad, and most importantly $\mathbb{T}$ is the torsion scalar and defined as 
\be
\mathbb{T}=-\mathcal{T}_{\mu}\mathcal{T}^{\mu}+\frac{1}{4}\mathcal{T}_{\alpha\beta\mu}\mathcal{T}^{\alpha\beta\mu}
+\frac{1}{2}\mathcal{T}_{\alpha\beta\mu}\mathcal{T}^{\beta\alpha\mu}~,
\ee
with $\mathcal{T}_{\mu}=\mathcal{T}^{\alpha}_{~~\mu\alpha}$ being the torsion vector.  
This action is diffeomorphism invariant and identical to the Einstein-Hilbert action up to a boundary term,
\be\label{graction}
S_g=-\frac{M_p^2}{2}\int d^4x \sqrt{-g}[R (e)+2\nabla_{\mu}\mathcal{T}^{\mu}]~,
\ee
where the curvature scalar $R(e)$ is defined by the Levi-Civita connection and considered as being fully constructed from the metric, and in turn from the tetrad. The covariant derivative $\nabla_{\mu}$ is also associated with the Levi-Civita connection. The boundary term does not affect the equation of motion, so the TEGR model is equivalent to GR. In addition, it can be also considered as a pure tetrad theory. The spin connection only contributes to the boundary term, so it represents pure gauge in the TEGR action (\ref{TGaction}), and in practice we may fix a spin connection (as long as it satisfies Eq. (\ref{omega})) and this does not affect the equation of motion. The simplest choice is making use of the Weitzenb\"{o}ck connection, $\omega^A_{~B\nu}=0$, which has been frequently adopted in the literature.
With the Weitzenb\"{o}ck connection, the torsion two form is simply expressed as 
\be\label{tor}
\mathcal{T}^{A}_{~~\mu\nu}=e^A_{~\rho}\mathcal{T}^{\rho}_{~~\mu\nu}=\partial_{\mu}e^{A}_{~\nu}-\partial_{\nu}e^{A}_{~\mu}~.
\ee
It deserves stressing that in a general TG theory fixing a spin connection usually means the broken of the local Lorentz symmetry, but it is not the case in the TEGR model (\ref{TGaction}) because the spin connection in this model only contributes the boundary term as shown in its equivalent form, Eq. (\ref{graction}). One can prove straightforwardly that,  when taking the Weitzenb\"{o}ck connection, the action (\ref{TGaction}) is unchanged under the local Lorentz transformation, $e^{A}_{~\mu}\rightarrow \Lambda^{A}_{~B}(x)e^{B}_{~\mu}$, up to a boundary term. 
However, for the modified TEGR models taking the Weitzenb\"{o}ck connection indeed breaks the local Lorentz symmetry, see Refs. \cite{PVtele1,PVtele2,PVtele3,PVtele4} for recent discussions. To avoid such explicit Lorentz violation in these cases, it is better to keep the general form (\ref{omega}) for the spin connection and treat both the tetrad $e^A_{~\mu}$ and the Lorentz matrix element $\Lambda^A_{~B}$ in Eq. (\ref{omega}) as fundamental variables.

Within the framework of TEGR, a gravitational baryogenesis model similar to that in Ref. \cite{Davoudiasl:2004gf} can be constructed by considering the derivative coupling of the torsion scalar to the baryon current, $\mathcal{L}_{in}\sim \partial_{\mu}\mathbb{T} J^{\mu}_B$. This gives rise to the baryon an effective chemical potential which is proportional to $\mathbb{\dot T}$. The standard cosmology is unchanged since TEGR is equivalent to GR\footnote{In gravitational baryogenesis models, the back reaction of the introduced derivative coupling to the spacetime can be ignored because in the radiation dominated epoch the baryon density is subdominant.}. The key point is that $\mathbb{T}=-R-2\nabla_{\mu}\mathcal{T}^{\mu}$ is different from $-R$ by the term $-2\nabla_{\mu}\mathcal{T}^{\mu}$, so that $\mathbb{ T}$ and its time derivative do not vanish in the radiation dominated epoch. This makes the effective chemical potential $\mu_b$ for the baryon non-vanishing, then is expected to generate a net baryon number according to Eq. (\ref{chemical}). 

In this section we consider the full action containing this derivative coupling
\be\label{actionfull}
S=\int d^{4}x ~\sqrt{-g}\left(\frac{M_p^2}{2}\mathbb{T}+\frac{1}{M_{\ast}^2}\,\partial_{\mu}\mathbb{T}J^{\mu}
\right)+S_{m}~.
\ee
Other matter and non-gravitational interactions including the baryon number non-conserving interactions are described by $S_{m}$. All the matter are assumed to couple to the metric (or to the tetrad) minimally besides the introduced derivative coupling. Because the derivative coupling depends on the torsion scalar which accounts for the gravity, this model is also classified to the modified TEGR model. 

The equations of motion follow from the variation of the action with respect to $e^{A}_{~\mu}$ and $\Lambda^{A}_{~B}$ separately:
\bea
(1-\frac{2\theta}{M_{\ast}^2M_p^2})G^{\mu\nu}+\frac{\theta}{M_{\ast}^2M_p^2}\mathbb{T}\,g^{\mu\nu}-\frac{2}{M_{\ast}^2M_p^2}S^{\mu\nu\sigma}\partial_{\sigma}\theta &=& \frac{1}{M_p^2}T^{\mu\nu}~,\label{Einstein}\\
 S^{[\mu\nu]\sigma}\partial_{\sigma}\theta &=& 0~,\label{Smunu}
\eea
where $\theta\equiv\nabla_{\mu}J^{\mu}$,  $G^{\mu\nu}$ is Einstein tensor,
$T^{\mu\nu}=-(2/\sqrt{-g})(\delta S_m/\delta g_{\mu\nu})$ is energy-momentum tensor of  matter, and $S^{\mu\rho\sigma}=(1/2)\mathcal{T}^{\mu\rho\sigma}-\mathcal{T}^{[\rho\sigma]\mu}+2g^{\mu[\rho}\mathcal{T}^{\sigma]}$ is the so-called superpotential in TG theory and antisymmetric under the interchange of its last two indices. 

Now we apply these equations to the spatially-flat Friedmann-Robertson-Walker (FRW) universe with $ds^2=dt^2-a^2(t) d\vec{x}^2$. Our purpose is to obtain the values of $\mathbb{T}$ and its time derivative. Given the FRW metric,
we can always choose the tetrad as $e^A_{~\mu}={\rm diag}\{1,a,a,a\}$. Then the spin connection will be solved through equations (\ref{Einstein}) and (\ref{Smunu}). We first consider the case in which the current in the derivative coupling is the baryon current $J^{\mu}=J^{\mu}_B$, so $\theta=\dot n_B+3Hn_B$ only depends on time but does not vanish during the baryogenesis process (in the radiation dominated epoch) due to the existence of baryon number violation. 
With these, the equations (\ref{Einstein}) and (\ref{Smunu}) require that $S^{[\mu\nu]0}=0$ and $S^{ij0}\propto \delta^{ij}$. These in turn give the following constraints on the spin connections:
\be\label{constraints}
\omega^a_{~bj}\delta^j_a\delta^b_i=0,~\omega^a_{~0i}\propto \delta^a_i~.
\ee
The next step is to find the Lorentz matrix elements $\Lambda$ to satisfy above constraints according to the expression of the spin connection (\ref{omega}). In the homogeneous and isotropic spacetime it is natural to consider the homogeneous $\Lambda$s as the solution to the equations. Indeed, if all the Lorentz matrix elements in Eq. (\ref{omega}) only rely on the time, then constraints in Eq. (\ref{constraints}) are satisfied automatically.  These considerations lead to the following result:
\be
\mathbb{T}=-R-2\nabla_{\mu}\mathcal{T}^{\mu}=-R-2(\partial_0+3H)(3H-\frac{1}{a}\omega^a_{~0i}\delta^i_a)-2\partial_i(\frac{1}{a}\omega^0_{~a0}\delta^a_i)=-6H^{2}~.
\ee
This result means that the torsion scalar does not vanish during the radiation dominated epoch and and its time derivative $\mathbb{\dot T}=-12H\dot H$ provides an effective chemical potential for baryons:
$\mu_b=\mathbb{\dot T}/M_{\ast}^2$, then induces the baryon number asymmetry:
\be
\frac{n_B}{s}=\frac{15}{2\pi^2}\frac{\mathbb{\dot T}}{g_{\ast s}M_{\ast}^2 T}=-\frac{90}{\pi^2}\frac{H\dot H}{g_{\ast s}M_{\ast}^2 T}=\frac{180}{\pi^2}\frac{H^3}{g_{\ast s}M_{\ast}^2 T}~,
\ee
at the last step we have used the relation $\dot H=-2H^2$ in the radiation dominated epoch. Again from the standard big bang cosmology it is well known that the Hubble rate in the radiation dominated epoch is $H\simeq g^{1/2}_{\ast} T^2/(3M_p)$, where $g_{\ast}$ counts the total degrees of freedom which contributes to the radiation density, it is equal to $g_{\ast s}$ at the very early universe. Hence the baryon number asymmetry is 
\be
\left. \frac{n_B}{s}\right |_{T_D}=\frac{20}{3\pi^2}g^{1/2}_{\ast} \frac{T_D^5}{M_p^3 M_{\ast}^2}\simeq 0.5 \times 10^{-54}\left(\frac{T_D}{\rm GeV}\right)^3\left(\frac{T_D}{M_{\ast}}\right)^2~.
\ee
At the second step of the above equation, we have considered $g_{\ast}= 106.75$ and $M_p\simeq 2.4\times 10^{18}$ {\rm GeV}. The decoupling temperature of baryon number non-conserving interaction is about $T_D\sim 100$ GeV \cite{Kuzmin:1985mm} and the cut-off scale $M_{\ast}$ should not be lower than it. With these considerations one can evaluate that the produced baryon number asymmetry is at most at the order $10^{-48}$. This is too small to be consistent with observations. 
This disappointing consequence has been also obtained in Ref. \cite{Oikonomou:2016jjh}, the authors then turned to the modified TEGR gravity, i.e., the $f(\mathbb{T})$ model, or the way of replacing the original derivative coupling by $\partial_{\mu}f(\mathbb{T})J^{\mu}_B$. Here we would like to point out that within the framework of TEGR model there is another way to circumvent this difficulty: gravitational leptogenesis. 

In the leptogenesis scenario \cite{lep1,lep2,lep3,lep4},  there should be $B-L$ violating processes at high energy scales, these can be realized by purely lepton number violations. In our gravitational leptogenesis model, in addition to the $B-L$ violating interactions ever in thermal equilibrium, we need also to identify the current $J^{\mu}$ in the action (\ref{actionfull}) with $J^{\mu}_{B-L}$. This means that the torsion scalar couples derivatively to the $B-L$ current in stead of the baryon current. Through similar calculations  we get the thermally produced $B-L$ asymmetry:
\be
\left. \frac{n_{B-L}}{s}\right |_{T_D}=\frac{20}{3\pi^2}g^{1/2}_{\ast} \frac{T_D^5}{M_p^3 M_{\ast}^2}\simeq 0.5 \times 10^{-54}\left(\frac{T_D}{\rm GeV}\right)^3\left(\frac{T_D}{M_{\ast}}\right)^2~,
\ee
where $T_D$ is the decoupling temperature of $B-L$ violating interaction and can be much higher than the electroweak scale. 
At later time the already produced $n_{B-L}/s$ asymmetry will not be washed out by the electroweak Sphaleron processes, which violate $B+L$ but conserve $B-L$ and are in thermal equilibrium when the temperature of the universe is at the range $100~ {\rm GeV} <T<10^{12}~ {\rm GeV}$ \cite{Kuzmin:1985mm}. Furthermore, the electroweak Sphaleron processes will convert partially the $B-L$ asymmetry to the baryon number and lepton number asymmetries respectively \cite{Buchmuller:2005eh}:
\be
\frac{n_B}{s}=c_s\frac{n_{B-L}}{s}~,~\frac{n_L}{s}=(c_s-1)\frac{n_{B-L}}{s}~,
\ee
where $c_s=(8N_f+4)/(22N_f+13)$ and $N_f$ is the number of generations. For the standard model $N_f=3$ and $c_s\simeq 0.35$, so the finally produced baryon number asymmetry in this model is
\be
\frac{n_{B}}{s}\simeq 0.18 \times 10^{-54}\left(\frac{T_D}{\rm GeV}\right)^3\left(\frac{T_D}{M_{\ast}}\right)^2~.
\ee
Numerically, if $T_D$ is close to $M_{\ast}$, the current observational result, $n_B/s\sim 8.7\times 10^{-11}$, requires the decoupling temperature $T_D$ should be $0.78\times 10^{15}$ GeV. We must point out that $T_D$ should be lower than the inflation scale, otherwise the produced asymmetry $n_{B-L}/s$ would be diluted during inflation. One may evaluate that the Hubble rate at $T_D$ is $H(T_D)\simeq g^{1/2}_{\ast} T_D^2/(3M_p)\sim 10^{12}$ GeV, it is lower than the energy scale of the inflation process in the single field inflation models.

\section{Gravitational leptogenesis within symmetric teleparallel gravity}

Now we turn to the gravitational leptogenesis model within the framework of symmetric teleparallel gravity (STG). This has not been discussed before. The STG theory can also be considered as a constrained metric-affine theory.  It is formulated in a spacetime with a metric $g_{\mu\nu}$ and an affine connection ${\Gamma^{\lambda}}_{\mu\nu}$, the latter leads to zero curvature and zero torsion,
  \be\label{noncurvature}
{\hat{R}^{\rho}}_{~\sigma\mu\nu}\equiv \partial_{\mu}{ \hat\Gamma^{\rho}}_{~\nu\sigma}-\partial_{\nu}{\hat{\Gamma}^{\rho}}_{~\mu\sigma}+{\hat{\Gamma}^{\rho}}_{~\mu\alpha}{\hat{\Gamma}^{\alpha}}_{~\nu\sigma}-{\hat{\Gamma}^{\rho}}_{~\nu\alpha}{\hat{\Gamma}^{\alpha}}_{~\mu\sigma}=0~, ~\mathcal{T}^{\rho}_{~~\mu\nu}=2\hat{\Gamma}^{\rho}_{[\mu\nu]}=0~.
 \ee
 With these constraints, the affine connection can be generally expressed as
 \begin{eqnarray} \label{gamma=}
	{\hat{\Gamma}^{\lambda}}_{~\mu \nu} (x) = \frac{\partial x^{\lambda}}{\partial y^{\beta}} \partial_{\mu}\partial_{\nu} y^{\beta}~,
\end{eqnarray}
with $y^{\beta}(x)$ being four functions. These functions defined a special coordinate system in which the affine connection vanishes. One may fix to the $y$-coordinate system to do the rest calculations. This is a gauge choice. In fact this ``coincident gauge" has been adopted extensively in the studies on STG theories in the literature. However, taking the coincident gauge will break the diffeomorphism invariance explicitly, which we try to avoid in this paper. So we will keep the general form (\ref{gamma=}) for the affine connection. 
The gravity in the STG theory is identical to the non-metricity. As usual the non-metricity tensor is defined as
\begin{eqnarray}\label{qdefinition}
	Q_{\alpha \mu \nu}\equiv \hat{\nabla}_{\alpha} g_{\mu \nu} = \partial_{\alpha} g_{\mu\nu} - {\hat{\Gamma}^{\lambda}}_{~\alpha\mu}g_{\lambda\nu} - {\hat{\Gamma}^{\lambda}}_{~\alpha\nu}g_{\mu\lambda}~,
\end{eqnarray}
which measures the failure of the affine connection to be metric-compatible. The STG Equivalent of GR (STGR) model has the following action,
\begin{eqnarray} \label{action Q}
	S_{g}=\frac{M_p^2}{2} \int d^{4}x \sqrt{-g}\mathbb{Q}\equiv \frac{M_p^2}{2} \int d^{4}x \; \sqrt{-g} \left[  \frac{1}{4}Q_{\alpha\mu\nu}Q^{\alpha\mu\nu} -\frac{1}{2}Q_{\alpha\mu\nu}Q^{\mu\nu\alpha}-\frac{1}{4}Q^{\alpha}Q_{\alpha}+\frac{1}{2}{\tilde{Q}}^{\alpha}Q_{\alpha}   \right] ~,
\end{eqnarray}
where $\mathbb{Q}$ is non-metricity scalar and the vectors $Q_{\alpha}, \tilde{Q}_{\alpha}$ are two different traces of the non-metricity tensor, i.e., $Q_{\alpha} = g^{\sigma\lambda}Q_{\alpha\sigma\lambda},~ \tilde{Q}_{\alpha} =  g^{\sigma\lambda}Q_{\sigma\alpha\lambda}$. 
In terms of the constraints (\ref{noncurvature}), one can easily find that $\mathbb{Q}=-R-\nabla_{\mu}(Q^{\mu} - \tilde{Q}^{\mu})$ so that the STGR action (\ref{action Q}) is equal to the Einstein-Hilbert action up to a boundary term,
\begin{equation}\label{action-g2}
	S_g=- \frac{M_p^2}{2}\int d^{4}x   \sqrt{-g}\left[  R+\nabla_{\mu}(Q^{\mu} - \tilde{Q}^{\mu}) \right] ~. 
\end{equation}
Hence, the STGR model is equivalent to GR.

By introducing the derivative coupling of the non-metricity scalar, the full action we consider is
\begin{eqnarray}\label{stgr}
	S=\int d^4 x \sqrt{-g} (\frac{M_p^2}{2}\mathbb{Q}+ \frac{1}{M_{\ast}^2}\partial_{\mu} \mathbb{Q} J^{\mu})+S_m~.
\end{eqnarray}
The equation of motion of via the variation with the metric is
\begin{eqnarray}
	(1-\frac{2\theta}{M_{\ast}^2M_p^2})G^{\mu\nu}  +\frac{\theta}{M_{\ast}^2M_p^2}\mathbb{Q} \, g^{\mu\nu} +\frac{1}{M_{\ast}^2M_p^2} P^{\alpha\mu\nu} \partial_{\alpha}\theta =\frac{1}{M^2_p} T^{\mu\nu}~,
\end{eqnarray}
where again we have $\theta\equiv \nabla_{\mu} J^{\mu}$ and is equal to $\dot n+3Hn$ in the FRW universe. The superpotential is defined as $P^{\alpha\mu\nu} = Q^{\alpha\mu\nu} - 2Q^{(\mu\nu)\alpha} -(Q^{\alpha} -\tilde{Q}^{\alpha})g^{\mu\nu} + g^{\alpha(\mu}Q^{\nu )}$, which is symmetric under the interchange of the last two indices. 
Besides the metric, the four functions $y^{\nu}$ from which the affine connection is constructed are independent variables in this model. Its corresponding equation of motion is 
\begin{eqnarray} \label{affine Eoms}
	\hat{\nabla}_{\alpha}\hat{\nabla}_{\mu} (\sqrt{-g}{P^{\alpha\mu}}_{\nu} \theta) =0~.
\end{eqnarray}

Similarly, our main purpose in this section is to find out the value of $\mathbb{Q}$ and its time derivative by solving above equations in the FRW universe. Since the space of FRW universe is homogeneous and isotropic, it has 
six spatial Killing vectors $\xi^{\mu}$. It is reasonable to require the affine connection is also uniformly distributed in the universe, so that its Lie derivatives along the Killing vectors vanish: $\mathcal{L}_{\xi} {\Gamma^{\lambda}}_{\mu\nu}=0$. With this symmetry requirement, the non-vanished components of the affine connection have the following forms \cite{Hohmann:2019fvf}:
\begin{eqnarray}
	{\hat{\Gamma}^{0}}_{~00} = \mathcal{K}_{1}(t),\, {\hat{\Gamma}^{0}}_{~11}={\hat{\Gamma}^{0}}_{~22}={\hat{\Gamma}^{0}}_{~33} =\mathcal{K}_{2}(t) ,\, {\hat{\Gamma}^{1}}_{~01}={\hat{\Gamma}^{2}}_{~02}={\hat{\Gamma}^{3}}_{~03}=\mathcal{K}_{3}(t)~,
\end{eqnarray}
where $\mathcal{K}_{1}(t), \mathcal{K}_{2}(t), \mathcal{K}_{3}(t)$ are three uniform functions. The zero curvature condition requires that
\begin{eqnarray}
	\mathcal{K}_{3} (\mathcal{K}_{1} - \mathcal{K}_3) - \dot{\mathcal{K}}_{3}  =0~, ~\mathcal{K}_{1}\mathcal{K}_{2} + \dot{\mathcal{K}}_{2} = 0~,~
	\mathcal{K}_{2}\mathcal{K}_{3} =0~.
\end{eqnarray}

Now we discuss the cases of $\mathcal{K}_{2}=0$ and  $\mathcal{K}_{3}=0$ separately. In the first case, $\mathcal{K}_{2}=0$, one can obtain the non-metricity scalar and the simple form of constraint equation (\ref{affine Eoms}) as follows,
\be
\mathbb{Q} = 3(-2H^2 +3H\mathcal{K}_{3}+\dot{\mathcal{K}}_{3})~,~\mathcal{K}_{3} (\ddot{\theta} + 3H\dot{\theta}) =0~. 
\ee
As mentioned before, $\theta=\dot n+3Hn$ does not vanish due to the non-conservation of corresponding quantum number. Its change with time depends on specific particle physics model on the quantum number violation. So in general, we cannot expect a vanishing $\ddot{\theta} + 3H\dot{\theta}$. Then the constraint equation (\ref{affine Eoms}) requires that $\mathcal{K}_{3}=0$ and we get the result $\mathbb{Q} = -6H^2$. 

In the second case $\mathcal{K}_{3}=0$, we get the non-metricity scalar and the form of constraint equation (\ref{affine Eoms}) as follows,
\begin{eqnarray}
	\mathbb{Q} = 3(-2H^2 +H\mathcal{K}_{2}/a^2+\dot{\mathcal{K}}_{2}/a^2)~,~ \mathcal{K}_{2}  (\ddot{\theta} - 2\mathcal{K}_{1}\dot\theta +H\dot{\theta})=0~.
\end{eqnarray}
Similarly we cannot expect $\ddot{\theta} - 2\mathcal{K}_{1}\dot\theta +H\dot{\theta}=0$ in general and we can only have $\mathcal{K}_{2} =0$. So we obtain the result $\mathbb{Q} = -6H^2$ again.

In all, we get $\mathbb{Q} = -6H^2$ for the model (\ref{stgr}) within the framework of STG theory. Next we have the same story about the baryon number asymmetry as that we have talked within the framework of TG theory in the previous section. If the current in the derivative coupling $(1/M_{\ast}^2)\partial_{\mu} \mathbb{Q} J^{\mu}$ is the baryon current $J^{\mu}_B$, the produced baryon number asymmetry is as low as $10^{-48}$ and cannot be consistent with observations. To have a way out of this problem, one may consider the modified STGR model, such as the $f(\mathbb{Q})$ model, or change the coupling $\partial_{\mu} \mathbb{Q} J_B^{\mu}$ to $\partial_{\mu} f(\mathbb{Q}) J_B^{\mu}$. In this paper we prefer the gravitational leptogenesis mechanism, i.e., besides assuming the existence of $B-L$ violating processes in the very early universe, the introduced derivative coupling of $\mathbb{Q}$ should be $(1/M_{\ast}^2)\partial_{\mu} \mathbb{Q} J_{B-L}^{\mu}$. 
Consequently we have the $B-L$ asymmetry
\be
\left. \frac{n_{B-L}}{s}\right |_{T_D}\simeq 0.5 \times 10^{-54}\left(\frac{T_D}{\rm GeV}\right)^3\left(\frac{T_D}{M_{\ast}}\right)^2~,
\ee
and a similar order asymmetry for baryons: $n_{B}/s\simeq 0.18 \times 10^{-54}(T_D/{\rm GeV})^3(T_D/{M_{\ast}})^2$.
 With the decoupling temperature of $B-L$ violating interactions $T_D\sim M_{\ast} \sim10^{15}$ GeV, the required baryon-to-entropy ratio $n_{B}/s\sim 10^{-10}$ can be obtained.

\section{conclusion}\label{conclusion}
In this paper we have studied the gravitational baryogenesis and leptogenesis models within the frameworks of teleparallel gravity (TG) and symmetric teleparallel gravity (STG). Both TG and STG theories can have models equivalent to GR, i.e., TEGR and STGR respectively, but account for gravitational phenomena from different viewpoints. By introducing the derivative couplings of the torsion scalar (in TEGR) or the non-metricity scalar (in STGR) to baryons, the baryon number asymmetry can be produced in thermal equilibrium. In the case of baryogenesis considered here, the produced baryon number asymmetry is too small to be consistent with observations. However, as we have shown, the leptogenesis scenario works for these cases.

{\it Acknowledgement}: This work is supported by NSFC under Grant No. 12075231, 11653002, 12047502 and 11947301.

{}


\begin{thebibliography}{}

\bibitem{Planck:2018vyg}
N.~Aghanim \textit{et al.} [Planck],
Astron. Astrophys. \textbf{641} (2020), A6
[erratum: Astron. Astrophys. \textbf{652} (2021), C4]
doi:10.1051/0004-6361/201833910
[arXiv:1807.06209 [astro-ph.CO]].

\bibitem{ParticleDataGroup:2020ssz}
P.~A.~Zyla \textit{et al.} [Particle Data Group],
PTEP \textbf{2020} (2020) no.8, 083C01
doi:10.1093/ptep/ptaa104

\bibitem{Sakharov:1967dj}
A.~D.~Sakharov,
Pisma Zh. Eksp. Teor. Fiz. \textbf{5} (1967), 32-35
doi:10.1070/PU1991v034n05ABEH002497

\bibitem{Cohen:1987vi}
A.~G.~Cohen and D.~B.~Kaplan,
Phys. Lett. B \textbf{199} (1987), 251-258
doi:10.1016/0370-2693(87)91369-4

\bibitem{Li:2001st}
M.~z.~Li, X.~l.~Wang, B.~Feng and X.~m.~Zhang,
Phys. Rev. D \textbf{65} (2002), 103511
doi:10.1103/PhysRevD.65.103511
[arXiv:hep-ph/0112069 [hep-ph]].

\bibitem{Li:2002wd}
M.~Li and X.~Zhang,
Phys. Lett. B \textbf{573} (2003), 20-26
doi:10.1016/j.physletb.2003.08.041
[arXiv:hep-ph/0209093 [hep-ph]].

\bibitem{DeFelice:2002ir}
A.~De Felice, S.~Nasri and M.~Trodden,
Phys. Rev. D \textbf{67} (2003), 043509
doi:10.1103/PhysRevD.67.043509
[arXiv:hep-ph/0207211 [hep-ph]].

\bibitem{Kolb:1990vq}
E.~W.~Kolb and M.~S.~Turner,
Front. Phys. \textbf{69} (1990), 1-547

\bibitem{Davoudiasl:2004gf}
H.~Davoudiasl, R.~Kitano, G.~D.~Kribs, H.~Murayama and P.~J.~Steinhardt,
Phys. Rev. Lett. \textbf{93} (2004), 201301
doi:10.1103/PhysRevLett.93.201301
[arXiv:hep-ph/0403019 [hep-ph]].

\bibitem{Shiromizu:2004cb}
T.~Shiromizu and K.~Koyama,
JCAP \textbf{07} (2004), 011
doi:10.1088/1475-7516/2004/07/011
[arXiv:hep-ph/0403231 [hep-ph]].


\bibitem{Li:2004hh}
H.~Li, M.~z.~Li and X.~m.~Zhang,
Phys. Rev. D \textbf{70} (2004), 047302
doi:10.1103/PhysRevD.70.047302
[arXiv:hep-ph/0403281 [hep-ph]].

\bibitem{Tele}
R. ~Aldrovandi and J.~G.~Pereira, {\it Teleparallel Gravity}, Vol. 173. Springer, 23 Dordrecht, (2013).

\bibitem{Nester:1998mp}
J.~M.~Nester and H.~J.~Yo,
Chin. J. Phys. \textbf{37} (1999), 113
[arXiv:gr-qc/9809049 [gr-qc]].

\bibitem{Oikonomou:2016jjh}
V.~K.~Oikonomou and E.~N.~Saridakis,
Phys. Rev. D \textbf{94} (2016) no.12, 124005
doi:10.1103/PhysRevD.94.124005
[arXiv:1607.08561 [gr-qc]].


\bibitem{Bhattacharjee:2020jfk}
S.~Bhattacharjee,
Phys. Dark Univ. \textbf{30} (2020), 100612
doi:10.1016/j.dark.2020.100612
[arXiv:2005.05534 [gr-qc]].

\bibitem{Azhar:2020coz}
N.~Azhar, A.~Jawad and S.~Rani,
Phys. Dark Univ. \textbf{30} (2020), 100724
doi:10.1016/j.dark.2020.100724
[arXiv:2009.13293 [gr-qc]].

\bibitem{Alexander:2004us}
S.~H.~S.~Alexander, M.~E.~Peskin and M.~M.~Sheikh-Jabbari,
Phys. Rev. Lett. \textbf{96} (2006), 081301
doi:10.1103/PhysRevLett.96.081301
[arXiv:hep-th/0403069 [hep-th]].

\bibitem{Bahamonde:2021gfp}
S.~Bahamonde, K.~F.~Dialektopoulos, C.~Escamilla-Rivera, G.~Farrugia, V.~Gakis, M.~Hendry, M.~Hohmann, J.~L.~Said, J.~Mifsud and E.~Di Valentino,
[arXiv:2106.13793 [gr-qc]].

\bibitem{PVtele1}
M.~Li, H.~Rao and D.~Zhao,
JCAP \textbf{11}, 023 (2020)
doi:10.1088/1475-7516/2020/11/023
[arXiv:2007.08038 [gr-qc]].

\bibitem{PVtele2}
M.~Li, H.~Rao and Y.~Tong,
[arXiv:2104.05917 [gr-qc]].

\bibitem{PVtele3}
H.~Rao,
[arXiv:2107.08597 [gr-qc]].

\bibitem{PVtele4}
M.~Hohmann and C.~Pfeifer,
Eur. Phys. J. C \textbf{81} (2021) no.4, 376
doi:10.1140/epjc/s10052-021-09165-x
[arXiv:2012.14423 [gr-qc]].


\bibitem{Kuzmin:1985mm}
V.~A.~Kuzmin, V.~A.~Rubakov and M.~E.~Shaposhnikov,
Phys. Lett. B \textbf{155} (1985), 36
doi:10.1016/0370-2693(85)91028-7

\bibitem{lep1}
M.~Fukugita and T.~Yanagida,
Phys. Lett. B \textbf{174} (1986), 45-47
doi:10.1016/0370-2693(86)91126-3

\bibitem{lep2}
P.~Langacker, R.~D.~Peccei and T.~Yanagida,
Mod. Phys. Lett. A \textbf{1} (1986), 541
doi:10.1142/S0217732386000683

\bibitem{lep3}
M.~A.~Luty,
Phys. Rev. D \textbf{45} (1992), 455-465
doi:10.1103/PhysRevD.45.455

\bibitem{lep4}
R.~N.~Mohapatra and X.~Zhang,
Phys. Rev. D \textbf{46} (1992), 5331-5336
doi:10.1103/PhysRevD.46.5331

\bibitem{Buchmuller:2005eh}
W.~Buchmuller, R.~D.~Peccei and T.~Yanagida,
Ann. Rev. Nucl. Part. Sci. \textbf{55} (2005), 311-355
doi:10.1146/annurev.nucl.55.090704.151558
[arXiv:hep-ph/0502169 [hep-ph]].


\bibitem{Hohmann:2019fvf}
M.~Hohmann,
Symmetry \textbf{12} (2020) no.3, 453
doi:10.3390/sym12030453
[arXiv:1912.12906 [math-ph]].


\end{thebibliography}
\end{document}